\documentclass[jgrga]{agutex}
\usepackage[ansinew]{inputenc}
\usepackage{bbm}
\usepackage{graphicx}
\usepackage{amsmath}
\authorrunninghead{WIENGARTEN ET AL.}

\titlerunninghead{INNER-HELIOSPHERIC MHD SIMULATIONS}

\authoraddr{T. Wiengarten,
Institut für Theoretische Physik IV, Ruhr-Universität Bochum, 44780 Bochum, Germany
(tow@tp4.rub.de)}

\authoraddr{J. Kleimann,
Institut für Theoretische Physik IV, Ruhr-Universität Bochum, 44780 Bochum, Germany
(jk@tp4.rub.de)}

\authoraddr{H. Fichtner,
Institut für Theoretische Physik IV, Ruhr-Universität Bochum, 44780 Bochum, Germany
(hf@tp4.rub.de)}

\authoraddr{R. Cameron,
Max-Planck-Institut für Sonnensystemforschung, 37191 Katlenburg-Lindau, Germany
(cameron@mps.mpg.de)}

\authoraddr{J. Jiang,
Max-Planck-Institut für Sonnensystemforschung, 37191 Katlenburg-Lindau, Germany
(jiang@mps.mpg.de)}

\authoraddr{R. Kissmann,
Institut für Astro und Teilchenphysik, Universität Innsbruck, A-6020 Innsbruck, Austria 
(Ralf.Kissmann@uibk.ac.at)}

\authoraddr{K. Scherer,
Institut für Theoretische Physik IV, Ruhr-Universität Bochum, 44780 Bochum, Germany
(kls@tp4.rub.de)}

\begin{document}

%
%

\title{MHD Simulation of the Inner-Heliospheric Magnetic Field}

%
%



\authors{T. Wiengarten\altaffilmark{1}, J. Kleimann\altaffilmark{1}, H. Fichtner\altaffilmark{1}, R. Cameron\altaffilmark{2}, J. Jiang\altaffilmark{2}, R. Kissmann\altaffilmark{3}, and K. Scherer\altaffilmark{1}}

\altaffiltext{1}{Institut für Theoretische Physik IV, Ruhr-Universität Bochum, 44780 Bochum, Germany}

\altaffiltext{2}{Max-Planck-Institut für Sonnensystemforschung, 37191 Katlenburg-Lindau, Germany}

\altaffiltext{3}{Institut für Astro- und Teilchenphysik, Universität Innsbruck, A-6020 Innsbruck, Austria}

%
%


\begin{abstract}
Maps of the radial magnetic field at a heliocentric distance of ten solar
radii are used as boundary conditions in the MHD code CRONOS to
simulate a 3D inner-heliospheric solar wind emanating from the rotating
Sun out to 1~AU. The input data for the magnetic field are the result of
solar surface flux transport modelling  
using observational data of sunspot groups coupled with a current–sheet source
surface model. Amongst several advancements, this allows for higher
angular resolution than that of comparable observational data from
synoptic magnetograms. The required initial conditions for the other
MHD quantities are obtained following an empirical approach 
using an inverse relation between flux tube expansion and radial solar wind speed. The computations are performed
for representative solar minimum and maximum conditions, and the corresponding state of the solar wind up to the Earth’s orbit is obtained. After a successful comparison
of the latter with observational data, they can be used to drive outer-heliospheric
models.
\end{abstract}

%
%

%

\begin{article}

\section{Introduction}
Since Parker's explanation of the expansion of the solar atmosphere as the solar wind \citep{Parker-1958},
its basic, long-term average features have been validated by direct in-situ as well as indirect remote
measurements. At a radial distance of a few solar radii, i.e.\ at the so-called source surface,
the heliospheric magnetic field (HMF) is essentially directed radially and `frozen in' into the radially 
expanding solar wind plasma. Because the magnetic field remains, due to an anchoring of the field line footpoints 
in the solar atmosphere, influenced by the solar rotation, the field lines form three-dimensional Archimedean
or so-called Parker spirals \citep{Parker-1958} if, in a first approximation, the Sun is considered as a rigid
rotator. 

For shorter periods, a more complex magnetic field structure was suggested by \citet{Fisk-1996}, who took into
account both the motion of magnetic footpoints on the solar surface and the existence of a non-vanishing latitudinal
field component. He was able to derive analytical expressions which generalize the Parker field correspondingly.
Subsequent analytical investigations \citep{Zurbuchen-etal-1997, Kobylinski-2001, Schwadron-2002, Schwadron-etal-2008}
have quantified the complex structure of this so-called Fisk-field and its deviations from the Parker field, and
numerical simulations \citep{Lionello-etal-2006} have revealed that the original choice of parameters by 
\citet{Fisk-1996} represented an extreme, which has recently been confirmed via a study of the consequences of the
Fisk-field for cosmic ray modulation \citep{Sternal-etal-2011}.

Several authors have suggested modifications of both fields \citep{Jokipii-Kota-1989, Smith-Bieber-1991, Schwadron-2002,
Burger-Hitge-2004, Burger-etal-2008, Schwadron-etal-2008}, which have been quantitatively compared in a study by
\citet{Scherer-etal-2010}. While very useful for many purposes, these analytical representations of the HMF cannot be
used to reproduce actual measurements in sufficient detail at any given time in the solar activity cycle.
For solar activity minima they are often too crude to catch small-scale HMF structures and for 
the HMF during maximum solar activity none of these analytical representations is valid. The 
field expressions suggested by \citet{Zurbuchen-etal-2004} for use during solar maximum are
a simple modification of a representation for solar minimum \citep{Zurbuchen-etal-1997} obtained 
by assuming a rather unlikely so-called Fisk angle, see \citet{Sternal-etal-2011}.

While, physically, the HMF originates in the Sun, it is, conceptually, customary to distinguish for modelling purposes
between the {\it coronal magnetic field} -- filling the region from the solar surface out to a spherical `heliobase'
\citep{Zhao-Hoeksema-2010} at several (tens of) solar radii -- and the HMF beyond. This concept is used in many modelling
attempts, which can be divided in two groups, potential field reconstructions and magnetohydrodynamical (MHD) models, 
see \citet{Riley-etal-2006}. The latter approach is computationally more expensive but can account for more physics,
direct time dependence and self-consistency. Examples for such MHD modelling of the coronal magnetic field are 
\citet{Usmanov-Goldstein-2003}, who computed a (tilted) axisymmetric solution for solar minimum conditions with the
assumption of a constant polytropic index, \citet{Cohen-etal-2007}, who extended this to a varying polytropic index 
and solar maximum conditions, or \citet{Lionello-etal-2009, Riley-etal-2011}, who employed phenomenological heating 
functions that are unrelated to the magnetic field direction. In more recent MHD models \citep[e.g.,][]{Nakamizo-etal-2009, 
Feng-etal-2012} improved heating and momentum source functions have been used and the two-region set-up has been dropped 
by exploiting advanced numerical grids providing high resolution close to the Sun as well as avoiding both coordinate-related
singularities and extreme cell size differences. For the construction of the heating functions these models are, however, 
implicitly using results from potential field reconstructions as, e.g., described in \citet{Feng-etal-2010}. While, in general,
MHD treatments should be preferred, all models published so far still have severe limitations like those indicated for the 
mentioned examples.

Not only computationally advantageous, but also with respect to their ability to resolve structures beyond those that can
be handled by current MHD models \citep{Riley-etal-2006}, are potential field models. These pure potential field 
source surface (PFSS) models, initially developed by \citet{Altschuler-Newkirk-1969} and \citet{Schatten-etal-1969} were
significantly refined by recognizing that the observed photospheric field should first be corrected for line-of-sight 
projection and then matched to the radial component of the potential field \citep{Wang-Sheeley-1992}, by explicitly 
taking into account additional sheet currents \citep{Zhao-Hoeksema-1995} resulting in the so-called current sheet source 
surface (CSSS) model, and by connecting such approaches to the solar wind expansion \citep{Arge-Pizzo-2000}, which resulted
in the so-called Wang-Sheeley-Arge (WSA) model, which is also used for the two-region set-up \citep{Pizzo-etal-2011a}. 

In the new model presented in this paper we follow a similar approach, i.e.~we consider a two-region model distinguishing 
between a coronal magnetic field region and that of the HMF which are separated by the heliobase as described above.
In difference and improvement of earlier approaches, however, the input data for the magnetic field are the result of solar
surface flux transport (SFT) modelling \citep{Jiang-etal-2010} using observational data of sunspot groups coupled
with the CSSS model by \citet{Zhao-Hoeksema-1995}. This approach allows for higher 
resolution than that of comparable observational data from synoptic magnetograms. The heliobase conditions for the
other MHD quantities are obtained following the empirical approach by \citet{Detman-etal-2006} and \citet{Detman-etal-2011}, who employed the WSA approach.

Besides the need for high-resolution models of the three-dimensional magnetized solar wind for studies of 
coronal mass ejections \citep[see, e.g.~the recent review by][]{Kleimann-2012}, magnetic clouds 
\citep[see, e.g.][and references therein]{Dalakishvili-etal-2011} or corotating interaction regions 
\citep[see, e.g.][]{Gosling-Pizzo-1999}, there is the need for improved boundary conditions at 1~AU for large-scale heliospheric 
models \citep[see, e.g.][]{Pogorelov-etal-2009}. Our new model is of particular interest 
in the context of long-term modelling of the heliosphere, because with the employed method based on sunspot 
data the heliobase conditions can be reconstructed backward in time for more than three centuries \citep[see the recent 
papers by][]{Jiang-etal-2011a, Jiang-etal-2011b}. 

\section{The Model}
\label{model}
In its basic setup, CRONOS solves the equations of ideal MHD in a one-fluid model, this being a justifiable approach for the solar wind plasma \citep[e.g.][]{van-der-Holst-etal-2005}. In its normalized form (see Table~\ref{tab_norm}) the set of equations to be solved reads
\begin{eqnarray}
  \partial_t \rho + \nabla \cdot (\rho {\bf v}) &=& 0    \label{conti} \\
  \partial_t (\rho {\bf v}) + \nabla \cdot \left[\rho {\bf v v} 
    +  (p + |{\bf B}|^2/2) \ \mathbbm{1} - {\bf B B} \right]
  &=& {\bf f} \label{momentum}\\
  \partial_t e + \nabla \cdot \left[ (e+p+ |{\bf B}|^2/2) \ {\bf v}
    - ({\bf v} \cdot {\bf B}) {\bf B} \right] &=& {\bf v}\cdot{\bf f} \label{Eq:energyintergration} \\
   \partial_t {\bf B} + \nabla \times {\bf E}  &=& 0 \label{induction}
\end{eqnarray}
where $\rho$ is the mass density, ${\bf v}$ is the velocity of a fluid element, ${\bf B}$ and ${\bf E}$ describe the electromagnetic field, $e$ is the total energy density, and $p$ is the scalar thermal pressure. ${\bf f}$ is the sum of the gravitational force density ${\bf f}_g = -\rho GM_\odot/r^2{\bf e}_r$ and additional force terms that may enter. Furthermore, $\mathbbm{1}$ denotes the unit tensor, and the dyadic product is used. So far the system is underdetermined, and to complete the set of equations we use the closure relations
\begin{eqnarray}
  e &=&
  \frac{\rho \ |{\bf v}|^2}{2} +
  \frac{       |{\bf B}|^2}{2} +
  \frac{p}{\gamma-1} \label{Eq:energy}\\
  {\bf E} + {\bf v} \times {\bf B} &=& 0 \\
  \nabla \cdot {\bf B} &=& 0 ~, \label{sole0}
\end{eqnarray}
where $\gamma$ is the adiabatic exponent.\\
While in the isothermal case ($\gamma=1$) the last term in (\ref{Eq:energy}) is dropped and equation (\ref{Eq:energyintergration}) becomes redundant and needs not enter the calculations, a temporally varying temperature requires using the full set of equations with $\gamma\ne1$. We follow \citet{Pomoell-etal-2011} in setting $\gamma=1.05$ to accelerate the solar wind. To model shock structures correctly, it would be necessary to work with adiabatic exponents as actually present in the solar wind, which are higher. To accomplish this, further following the approach of \citet{Pomoell-etal-2011} we would take the so far obtained solution as solar wind background input for a new simulation with introduced shocks. However, the model presented here does not involve shock structures. Nevertheless, extending our model in this direction can be a subject for future work. 

\begin{table}[ht]
  \caption{\label{tab_norm} \em
    \it Summary of normalized quantities and normalization constants. The choice made here takes length in solar radii, and number density and magnetic induction according to typical values in the solar photosphere.}
	\begin{center}
  \begin{tabular}{c|rcl}
    \hline
    physical quantity &
    \multicolumn{3}{c}{\rule[-3mm]{0mm}{8mm}
      normalization constant} \\
    \hline
    \hline
    length $L$ & $L_0$ &:=&
    $6.96 \cdot 10^8$ m \\
    number density $n$ & $n_0$ &:=&
    $1.00 \cdot 10^{14}$ m$^{-3}$ \\
    magn. induction $B$ & $B_0$ &:=&
    $1.00 \cdot 10^{-4}$ T \\
    \hline
    mass density $\rho$ & $\rho_0 = m_{\rm p} n_0$ &=&
    $1.67 \cdot 10^{-13}$ kg m$^{-3}$ \\
    mass $m$ & $m_0 = m_{\rm p} n_0 L_0^3$ &=&
    $5.63 \cdot 10^{13}$ kg \\
    velocity $v$ & $v_0 = \frac{B_0}{\sqrt{\mu_0 \rho_0}}$ &=&
    $2.18\cdot10^5$ m s$^{-1}$\\
    acceleration $g$ & $g_0 = v_0^2 / L_0$ &=&
    $6.84\cdot10^1$ m s$^{-2}$\\
    time $t$ & $t_0 = L_0 / v_0$ &=&
    $3.19\cdot10^3$ s\\
    energy density $e$ & $e_0 = m_{\rm p} n_0 v_0^2$ &=&
    $7.96\cdot10^{-3}$ J  m$^{-3}$ \\
    current density $J$ & $J_0 = \sqrt{\frac{\rho_0}{\mu_0}}\frac{v_0}{L_0}$ &=&
    $1.14 \cdot 10^{-7}$ A  m$^{-2}$ \\
    gas pressure $p$ & $p_0 = m_{\rm p} n_0 v_0^2$ &=&
    $7.96 \cdot 10^{-3}$ Pa \\
    temperature $T$ & $T_0 = \frac{m_{\rm p} v_0^2}{2k_B}$ &=&
    $2.88 \cdot 10^{6}$ K \\
  \end{tabular}
  \end{center}
  \label{t:norm}
\end{table}
 
We use  spherical coordinates ($r$,$\vartheta$,$\varphi$) with the origin at the center of the Sun. Thus, $r$ is the heliocentric radial distance, $\vartheta\in[0,\pi]$ is the polar angle (with the north pole corresponding to $\vartheta=0$) and $\varphi\in[0,2\pi]$ is the azimuthal angle. The reference point for $\varphi$ is as follows: For the test cases (see Section \ref{WD}) azimuthal symmetry is used so that results are the same for all $\varphi$, and it is not necessary to define a reference point. The observationally based data in Section~\ref{observational} is given in Carrington longitudes and the position of the Earth serves as a reference point. It depends on the time at which the data were taken. This will be described in Section~\ref{inputdata} after explaining how the data were obtained.\\
Neglecting differential rotation, solar rotation can be described via an angular frequency
\begin{equation}
\Omega = 14.71 °/{\rm d}
\label{omega}
\end{equation} 
according to \citet{Snodgrass-Ulrich-1990}, which corresponds to the sidereal rotation period of 24.47 days. Accordingly, the Sun's rotation axis' tilt of about 7.25° relative to the axis of the Earth's orbit is also not taken into account.\\
Implementing solar rotation into the model can be done in both the solar wind plasma rest-frame and the frame co-rotating with the Sun. The first choice introduces azimuthal components of ${\bf B}$ and ${\bf v}$ by applying respective boundary conditions while solving the original set of equations (\ref{conti})-(\ref{sole0}). The latter choice cannot be implemented by boundary conditions only, but requires the introduction of ficititious force terms that occur in rotating (and thus non-inertial) frames of reference, namely the Coriolis force, the centrifugal force, and the Euler force:
\begin{equation}
{\bf f}_{cor} = - 2\rho{\bf\Omega}\times{\bf v} -  \rho{\bf\Omega}\times({\bf\Omega}\times{\bf r}) - \rho\frac{d\bf\Omega}{dt}\times{\bf r}~.
\label{fictitious1}
\end{equation}
In the chosen spherical coordinate system the angular velocity is directed along the $z$ axis of the corresponding Cartesian coordinate system, coinciding with the axis of rotation of the Sun: ${\bf\Omega} = \Omega{\bf e}_z$. Solar rotation is assumed constant in time ($\Omega\neq\Omega(t)$) so that the Euler force term vanishes. With the Cartesian unit vector transformed to the spherical system this finally gives 
\begin{eqnarray}
{\bf f}_{cor} =\rho\big[& &\left( 2\Omega\sin\vartheta \ v_{\varphi} + r\Omega^2\sin^2\vartheta\right){\bf e}_r \nonumber \\
						    				&+&\left( 2\Omega\cos\vartheta \ v_{\varphi} + r\Omega^2\sin\vartheta\cos\vartheta\right){\bf e}_{\vartheta} \nonumber \\
			 			    				&-&\left( 2\Omega(\cos\vartheta \ v_{\vartheta}+\sin\vartheta \ v_{r})\right){\bf e}_{\varphi}\big]~.
		 \label{F_cor}
\end{eqnarray}

\section{Model Validation: Weber \& Davis Model}\label{WD}
\subsection{Analytical formulation}
Mandatory tests have been performed to check the correct implementation of our model. First, Parker's solar wind model was reconstructed, using a respective isothermal model. It could be shown that implementations in both frames of reference produced correct results of the typical radial velocity profile, and magnetic field lines bent to spirals resulting from solar rotation were obtained. A more thorough comparison with analytical results, however, can be made with the more complex Weber \& Davis model \citep{Weber-Davis-1967} (hereafter WD): Parker's assumption of frozen-in field lines is not valid close to the Sun since there the kinetic energy density is smaller than the magnetic energy density. Consequently, it can be assumed that the plasma flow follows the magnetic field lines instead well inside the Alfvénic radius $r_A$. This co-rotation should weaken when approaching $r_A$, and for the limiting case $r \gg r_A$ Parker's model ($v_\varphi \propto r^{-1}$) should be a good approximation.\\
Assuming an azimuthal component of the flow velocity and solving a steady-state model with axial symmetry in the equatorial plane, the final equations of WD for the azimuthal components of velocity and magnetic induction in the solar wind plasma frame of reference read
\begin{equation}
	v_{\varphi}(r) = \Omega_{WD} r\frac{1-v_r/v(r_A)}{1-M_A^2}
	\label{vrot}
\end{equation}
\begin{equation}
	B_{\varphi}(r) = -B_r\frac{\Omega_{WD} r}{v(r_A)}\frac{1-(r/r_A)^2}{1-M_A^2}~.
	\label{brot}
\end{equation}
where $v(r_A)$ and $M_A$ denote the radial velocity at the Alfvénic critical point and the radial Alfvén Mach number, respectively. Furthermore, $B_r=B_0/r^2$ as follows from the solenoidality condition. It is important to note that, here, $\Omega_{WD}$ is, according to WD, the "angular velocity of the roots of the lines of force in the Sun," whose value is not easily accessible. It can be shown \citep{Barker-Marlborough-1982, MacGregor-Pizzo-1983} that $\Omega_{WD}$ is related to the observed angular velocity of the photosphere $\Omega$ (as in equation (\ref{omega})) via
\begin{equation}
\Omega_{WD} = (1+f)\Omega
\label{WDBM}
\end{equation}
with 
\begin{equation}
f = \frac{v_{0,r}|B_{0,\varphi}|}{v_{0,\varphi}B_{0,r}}
\end{equation}
where the quantities with zero-subscript indicate the respective values at the solar photosphere that are easily accessible in a simulation starting at $r=R_\odot$.\\
This derivation can be adopted for the co-rotating frame of reference as well, taking the initial value for the azimuthal velocity $v_{\varphi,co}(R_\odot) = 0$, and extending the equation of motion by the fictitious force terms (in the ecliptic plane). The final result is
\begin{eqnarray}
	v_{\varphi,co}(r) = \frac{1}{M_A^2-1} &\Bigg[&v_{0,r}\frac{R_\odot}{r}\frac{B_{0,\varphi}}{B_r(r)}\left(1-M_A^2\frac{B_r(r)}{B_r(r_A)}\right) \nonumber \\
																				&&- M_A^2\Omega r\left(1-\frac{r_A^2}{r^2}\right)\Bigg]
\label{vcor}
\end{eqnarray}
and
\begin{equation}
	B_{\varphi}(r) = \frac{(R_\odot/r)v_{0,r}B_{0,\varphi} + v_{\varphi}(r)B_r(r)}{v_r(r)}~.
\label{bcor}
\end{equation}

\subsection{Numerical simulations}
The analytic results shown above can be compared to the numerical results obtained from respective simulations in the co-rotating and the rest frame. In both cases the simulation box is restricted to extend from the Sun to ten solar radii and the number of grid points in radial direction is chosen to be $N_r = 400$, yielding a cell size of $\Delta r = 9R_\odot / 400 \approx 0.02R_\odot$. Since the WD solution is valid for the ecliptic plane only, $\vartheta$ is restricted to the interval $[0.4\pi,0.6\pi]$ and a resolution of $N_\vartheta = 41$, using an odd number, gives a cell centered at $\vartheta=\pi/2$ and a cell size of $\Delta\vartheta \approx 0.005\pi$. The complete $\varphi$ interval can be covered with a single layer of cells due to the symmetry.\\
For the implementation of boundary conditions, CRONOS makes use of so-called ghost cells that are extensions of the actual grid at its boundaries. $\vartheta$ boundaries are set to "reflecting", which basically ensures conservation of both mass- and magnetic flux, but influences the adjacent cells. This is compensated by a modest resolution so that there is no effect on the equatorial cells. Periodic boundary conditions are used for $\varphi$. The outer radial boundary condition is set to "outflow" and is self-explanatory. The inner radial boundary condition depends on the respective quantity and either fixes it at $r=R_\odot$ or extrapolates inwards into the boundaries ghost cells: For the initial number density, $n(r) = n_0(R_\odot/r)^3$ is fixed. Radial velocity is initially set to $v_r(r)=C\cdot r$ with an arbitrary constant $C=0.2$ so that the steady-state equation of continuity is fulfilled. For this flux conserving inward extrapolation is applied. The radial magnetic field is initialized as $B_r(r)=B_0(R_\odot/r)^2$. This is in accordance with the solenoidality condition since latitudinal components are set to zero ( $B_\vartheta=0$, $v_\vartheta=0$) and the azimuthal initialization (see below) yields $B_\varphi\neq B_\varphi(\varphi)$. \\
The azimuthal components of ${\bf B}$ and ${\bf v}$ depend on the frame of reference. For the rest frame, $v_\varphi(r) = \Omega r$, while in the corotating frame, $v_{\varphi,co}(r)=0$ initially, both being fixed at the inner boundary. The azimuthal magnetic field component is actually not initialized (i.e. set to zero), but the computed values at the innermost cells during runtime is extrapolated into the inner boundaries ghost cells according to 
\begin{equation}
B_\varphi(r\le R_\odot) = [B_\varphi(R_\odot + \Delta r)/B_r(R_\odot + \Delta r)]\cdot B_r(r) ~. \label{eq:bphi_test}
\end{equation}
This can be done for both reference frames because the magnetic field is invariant under the respective transformation.\\

\subsection{Comparison}
\begin{figure}
\noindent
\includegraphics[width=0.46\textwidth]{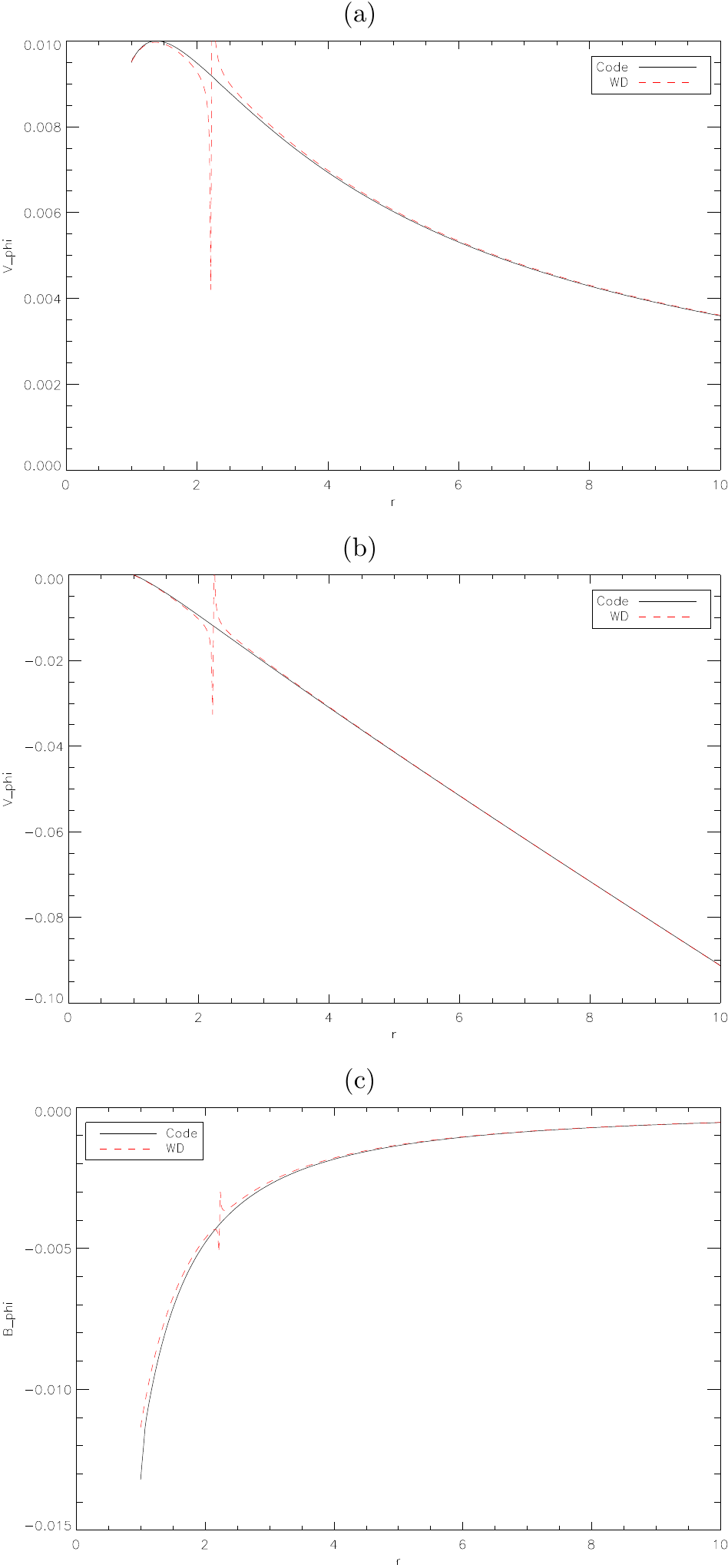}
\caption{\it (a) The azimuthal velocity (rest frame) exhibits co-rotation up to the Alfvénic critical point and afterwards decreases as $1/r$. It matches the analytic reference solution except at the Alfvénic critical point that, analytically, is a removable singularity, which cannot be resolved here. (b) Azimuthal velocity (co-rotating frame) where co-rotation cannot be seen as clearly as for the rest frame because the scale is ten times coarser. (c) The azimuthal magnetic field matches the analytic solution only for $r>4R_\odot$ because of the inner radial boundary condition (see text).}
\label{fig:WD}
\end{figure}

Figure \ref{fig:WD} shows the results for the azimuthal components for the rest frame and the co-rotating frame: In the rest frame (a) $v_\varphi$ exhibits the expected behavior as for \mbox{$R_\odot<r<1.5R_\odot$} a decreasing co-rotation is found. In the region near the Alfvénic critical radius at $r\approx2.2R_\odot$ the influence of the magnetic field is still evident: there is not yet an $1/r$ behavior as it develops for even larger $r$ in Parker's model. The WD solution (dashed line) matches the code's solution except near the Alfvénic point, which numerically cannot be exactly resolved due to a singularity that, analytically, is a removable one. For the co-rotating frame of reference (b) the code also reproduces the analytic solution. The co-rotating behavior of the plasma near the solar surface is only slightly recognizable because the $y$-axis scale is ten times coarser.\\
The azimuthal magnetic field (c) is also very similar to that of the WD solution, and for $r>4R_\odot$ they overlap. The discrepancy for smaller $r$ can be explained by the inner boundary condition that extrapolates the ratio between azimuthal and radial magnetic field component into the boundary. This gives higher absolute values in the boundary because $B_r$ has an $1/r^2$ behavior while for $B_\varphi$ it is an $1/r$ behavior. These higher absolute values are taken into the simulation box but their influence vanishes as $r$ increases. Consequently, a refinement of the boundary condition is not essential to get correct results for larger $r$.\\
In summary the code reproduces the analytic solutions of WD in both frames of reference while boundary conditions can lead to small deviations at the simulation box's boundaries if extrapolation is applied. 

\section{Observationally Based Inner Boundary Condition} \label{lindau}
\label{observational}
After successfully reproducing analytic results, we now use observationally based input data, i.e.~magnetic field distributions at ten solar radii, which are the result of a solar surface flux transport model coupled with an extrapolation of the heliospheric field that uses observational data of sunspot groups \citep{Jiang-etal-2010}. While this is the key input to our code, it is, however, necessary to provide the code with initial conditions for all MHD variables. For this purpose the approach by \citet{Detman-etal-2006,Detman-etal-2011} is followed, who use the empirical inverse relation between flux tube expansion factor and solar wind speed to determine the latter.

\subsection{Radial magnetic field}\label{inputdata}
\begin{figure*}
\noindent
\includegraphics[width=\textwidth]{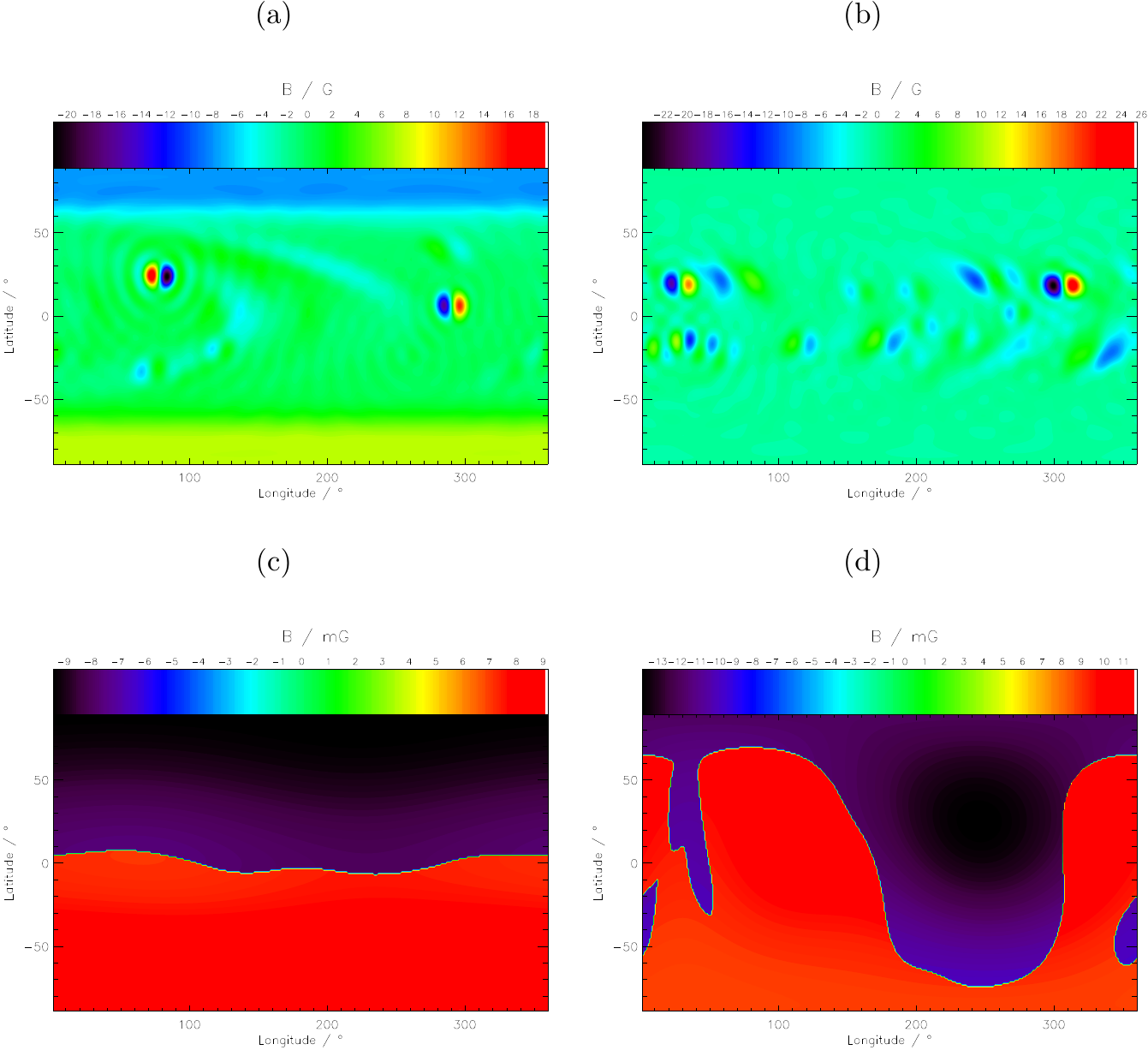}
\caption{\it Maps of the magnetic field strength at the solar surface (top row) and the source surface (bottom row) at $R_{ss} = 10R_\odot$ at solar minimum (in 1987.2, left) and solar maximum (in 2000.5, right) according to \citet{Jiang-etal-2010}.}
\label{fig:Rss}
\end{figure*}

As input data for our modelling, maps of the radial magnetic field at ten solar radii will be used. Their complete derivation is described in \citet{Jiang-etal-2010} and \citet{Schuessler-Baumann-2006}, but since it is the key input to the model presented here, in the following it is summarized how the input data was generated. The idea is to use sunspot group records as input to a solar surface flux transport (SFT) model yielding the magnetic flux at the solar surface, which is then extrapolated to ten solar radii by means of the so-called current sheet source surface (CSSS) model: The USAF/NOAA sunspot group records date back to 1976 and provide the basis of magnetic flux input, because sunspots are associated with emerging bipolar magnetic regions (BMRs). This allows for a model of the whole photosphere, whereas in observations, the back of the Sun cannot be seen. Along with flux cancellation and transport by surface flows the SFT model describes the formation of the magnetic flux distribution at the solar surface, which then is subject to latitudinal differential rotation, meridional flow, and turbulent diffusion due to granulation and supergranulation, so that all the relevant physical processes are implemented. Additionally, computation parameters for the magnetic flux are calibrated to match observed values.\\
In a next step the HMF has to be determined. The most widely used approach to achieve this is the potential field source surface (PFSS) model \citep{Schatten-etal-1969,Altschuler-Newkirk-1969}, which solves Laplace's equation between the photosphere and the so-called source surface at $r=R_{ss}$ at which the field is forced to be radial. \it Ulysses\rm\ data, however, motivated an extension of the PFSS model to explicitly take into account the heliospheric current sheet (HCS), yielding the CSSS model \citep{Zhao-Hoeksema-1995,Zhao-etal-2002}. As well as adding horizontal volume currents, the CSSS model includes the effect of sheet currents by introducing another spherical surface, the so-called cusp surface at $r=R_{cs}=1.55 R_{\odot}<R_{ss}=10 R_{\odot}$, within which the field contains only horizontal volume currents with a characteristic length scale of $0.2 R_{\odot}$. This corresponds to $a=0.2$ in the model of \citet{Bogdan-Low-1986}. Between the cusp surface and the source surface the field is configured by current sheets so that all field lines passing through the cusp surface reach the source surface. The field lines are assumed to be radial at the source surface. In our case the source surface is located at $R_{ss} = 10R_\odot$ and the results of the coupled SFT/CSSS model will be used as input to the model described here.\\
The question as to why the \it photospheric\rm\ magnetic field distributions are not used to run our MHD model, therefore starting at the Sun, can be answered as follows: First, an MHD code requires high spatial resolution near the Sun, so that even on massively parallel architectures it is time consuming. Additionally, such efforts are known to encounter difficult boundary condition related problems, because the inner radial boundary lies within the Alfvénic critical radius \citep{Nakagawa-1981a,Nakagawa-1981b}. Second, source surface models are easy to implement and require modest computer resources. The MHD approach as well as the source surface models are both simplifications of reality. Even if MHD models incorporate more physics, they cannot claim to cover everything. Furthermore, comparisons between the MHD and PFSS models have been performed by \citet{Riley-etal-2006} and their "results endorse the PFSS approach, under the right conditions and with appropriate caveats [...]." Given that they clearly acknowledge the significance of the various refinements of the PFSS model (e.g.~current sheets, non-spherical source surface), one should consider their finding as support not only for the usefulness of PFSS models but rather of static extrapolation models in general. Here, the input data has been derived using the augmented CSSS model, as well as a sophisticated SFT model so that it can be assumed that for the general survey presented here the magnetic field distributions at ten solar radii are a very good input.\\
One of the main advantages of the CSSS model is that, as discussed in \citet{Schuessler-Baumann-2006}, it yields an unsigned field strength at and beyond the source surface which is only very weakly latitude dependent. The PFSS model, by contrast has a strong latitudinal variability which is in conflict with the Ulysses spacecraft observations. This difference in the latitudinal distribution directly affects the expansion of field lines from the photosphere to the source surface, which is important in our modelling of the plasma properties of the inner heliosphere.
Hence for our purposes the CSSS model is preferred to that of the PFSS model.\\
In comparison with synoptic magnetograms often used as input data to models such as ours, our input data has some advantages since it allows for higher angular resolution and the whole photosphere can be modelled whereas in observations only part of the Sun is visible.\\
Even though a whole time series as output of these models exists, we will not make use of all of them in this case study, but reserve it for future work. Instead, the focus lies on two maps at different times in the solar cycle, i.e.~one at solar minimum (1987.2) and one at solar maximum (2000.5). The respective radial magnetic field maps are shown in Figure \ref{fig:Rss}. All figures are the same (apart from color scaling) as in Figure 4 (a)-(b) and (e)-(f) of \citet{Jiang-etal-2010}. At solar minimum (left panel) the Sun is relatively quiet, and, accordingly, the solar surface (top row) displays few to no sunspots and thus BMRs, while the overall structure is dipole-like. At the source surface (bottom row) the HCS is rather flat and the magnetic field reflects the dipole-like structure of the coronal magnetic field becoming radial at $R_{ss}$. However, the overall magnetic field strength is rather homogeneous. The latter is also true at solar maximum, but the HCS shows strong excursions to high latitudes, and even additional current sheets may occur. This is due to the large number of BMRs at the solar surface that are more prominent than the dipole structure.\\
The maps cover the full longitudinal and latitudinal intervals except for the poles (i.e.~$\vartheta\in[1°,179°]$) and have a resolution of $1°\times1°$. The latitudinal coordinate was described in Section \ref{model}. For the longitudinal coordinate Carrington longitudes are used that, together with a given time at which the data (which is a snapshot) was taken, define a reference system. It is, however, instructive to demonstrate how the position of the Earth can be found from the given time of the snapshot: The Carrington rotation system of reference is based on the synodic solar rotation rate as viewed from the Earth, giving the time required for one rotation $T_{Carr} = 27.2753$~days. At the beginning of a Carrington rotation the position of the Earth is at 360° with values decreasing to 0° towards the next rotation. The dates at which a new rotation commences can be found in respective tables (e.g.~{\tt http://alpo-astronomy.org/solar/rotn\_nos.html}). Taking the example of the dataset from $t_{snap} = 1987.2$, the respective Carrington rotation number is 1786 commencing at $t_{CR}$ = 1987.15727. The difference in time can be used to calculate the longitude of the Earth according to

\begin{equation}
\varphi_{Earth} = 360° - 360°\cdot\frac{t_{snap}-t_{CR}}{T_{Carr}}
\end{equation}
which puts the position of the Earth at respective values $\varphi_{1987.2} = 154°$ and $\varphi_{2000.5} = 114°$ (with Carrington rotation 1964 commencing at 2000.44899).

\subsection{Remaining MHD quantities}
The simulation requires initial values for the full set of the MHD variables, while so far only the radial magnetic field component is known from the previously described data set. One possible approach to derive the remaining quantities from the radial magnetic field data at the inner radial grid boundary is described in \citet{Detman-etal-2006,Detman-etal-2011} and shall be briefly summarized:\\
As input data a sequence of photospheric magnetic maps composed of daily magnetograms is used. A source surface current sheet model is used to extrapolate the HMF, which is then fed into an MHD code with an inner radial grid boundary at $R_{gb} = 0.1~{\rm AU} = 21.5R_\odot$. It is not possible to start the MHD simulations at the source surface most of the time. Instead, the Alfvénic critical point must be located outside the grid, which ensures that the plasma flow speed is higher than any characteristic plasma speed (specifically the fast mode wave speed), so that there cannot be any flow of information back into the boundary. The lower radial grid boundary is thus set to 0.1~AU, where this condition is met. It is then required to develop an interface to translate the input data from the source surface model to the lower radial grid boundary, which gives maps of all the necessary MHD variables there. This can only be achieved by {\it empirical} formulas, because it has to mimic the main plasma physics in the corona, which is not incorporated in source surface models. The resulting set of equations contains a number of adjustable parameters that can be tuned to give best fits to observational data at Earth. This is an ongoing process and will go hand in hand with advancements in other areas to further augment these kinds of models. The following set of equations is taken from \citet{Detman-etal-2006}, but has been adapted: An augmented formula for the radial velocity is taken from \citet{Detman-etal-2011}, which additionally uses the footpoint distance $d_{FP}$ to the nearest coronal hole boundary because using both footpoint distance and expansion factor $f_s$ gives better results than either one alone \citep{Arge-etal-2003}, which was found to be true for the data used here as well. Furthermore, the azimuthal velocity is transformed to the co-rotating reference frame. 
The following set of equations is to be understood as to give initial values at the inner radial grid boundary at $R_{gb}$, while initialisitation at larger radii will be adressed subsequently:
\begin{eqnarray}
v_r &=& v_{min} + \frac{v_{del}}{f_{s}^{V_{exp}}} + \alpha\cdot\left(\frac{d_{FP}}{f_{s}^{V_{exp}/2}} - \beta\right) \label{det:vr2}\\
\rho &=& F_{mass}/v_r \label{det:rho}\\
T_p &=& \frac{p_{tot} - {\bf B}^2/2}{\rho} \label{det:T}\\
B_r &=& b_{scale}\cdot B_{ss} \label{det:Br} \\
v_\vartheta &=& 0 \label{det:vtet} \\
B_\vartheta &=& 0 \label{det:Btet} \\
v_\varphi &=& \Omega R_\varphi\sin(\vartheta) - \Omega R_{gb}\sin(\vartheta) \label{det:vphi} \\
B_\varphi &=& (B_r/v_r)v_\varphi ~. \label{det:Bphi} 
\end{eqnarray} 
\\
Before describing the parameters in and the implementation of these formulas, it should be mentioned that there are differences between this model and \citet{Detman-etal-2006, Detman-etal-2011}. Most importantly, their extrapolation scheme puts the source surface at smaller radial distances, and the extrapolation method is not the same as the one that generated the input data used here. Thus, their parameter tuning might not be optimal for our purpose, but repeating this process for the data used here would go beyond the scope of this work.\\

\begin{figure}
\noindent
\includegraphics[width=0.5\textwidth]{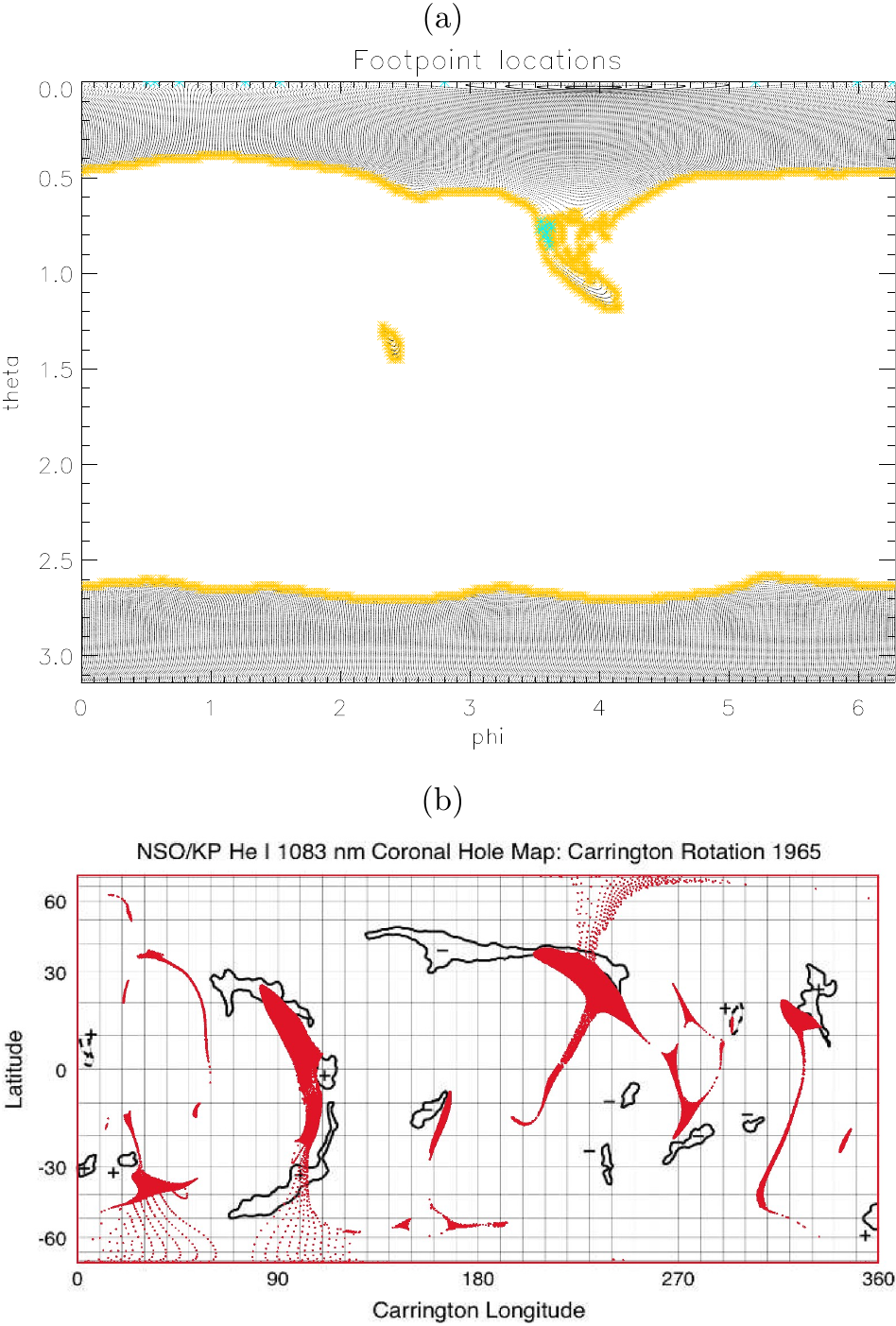}
\caption{\it Location of footpoints of open magnetic field lines in the photosphere (black/red crosses) for minimum (a) and maximum (b) conditions. In (a) yellow dots indicate coronal hole boundaries and cyan dots indicate field lines with the smallest expansion factors and thus result in the highest velocities. In (b) a comparison with coronal holes inferred from National Solar Observatory He maps (black contours) is superimposed. The indicated polarities were also found to be in good agreement with our data.}
\label{fig:footpoints}
\end{figure}

\begin{figure}
\noindent
\begin{center}
\includegraphics[width=0.5\textwidth]{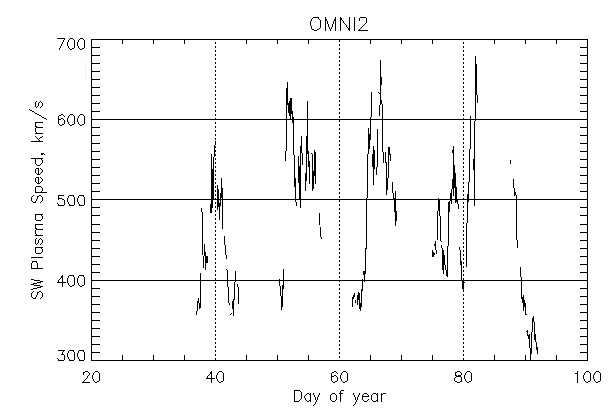}
\end{center}
\caption{\it OMNI solar wind speed data around 1987.2 measured with the \it IMP 8\rm\ spacecraft.}
\label{fig:omni1987}
\end{figure}

\begin{figure*}
\noindent
\includegraphics[width=\textwidth]{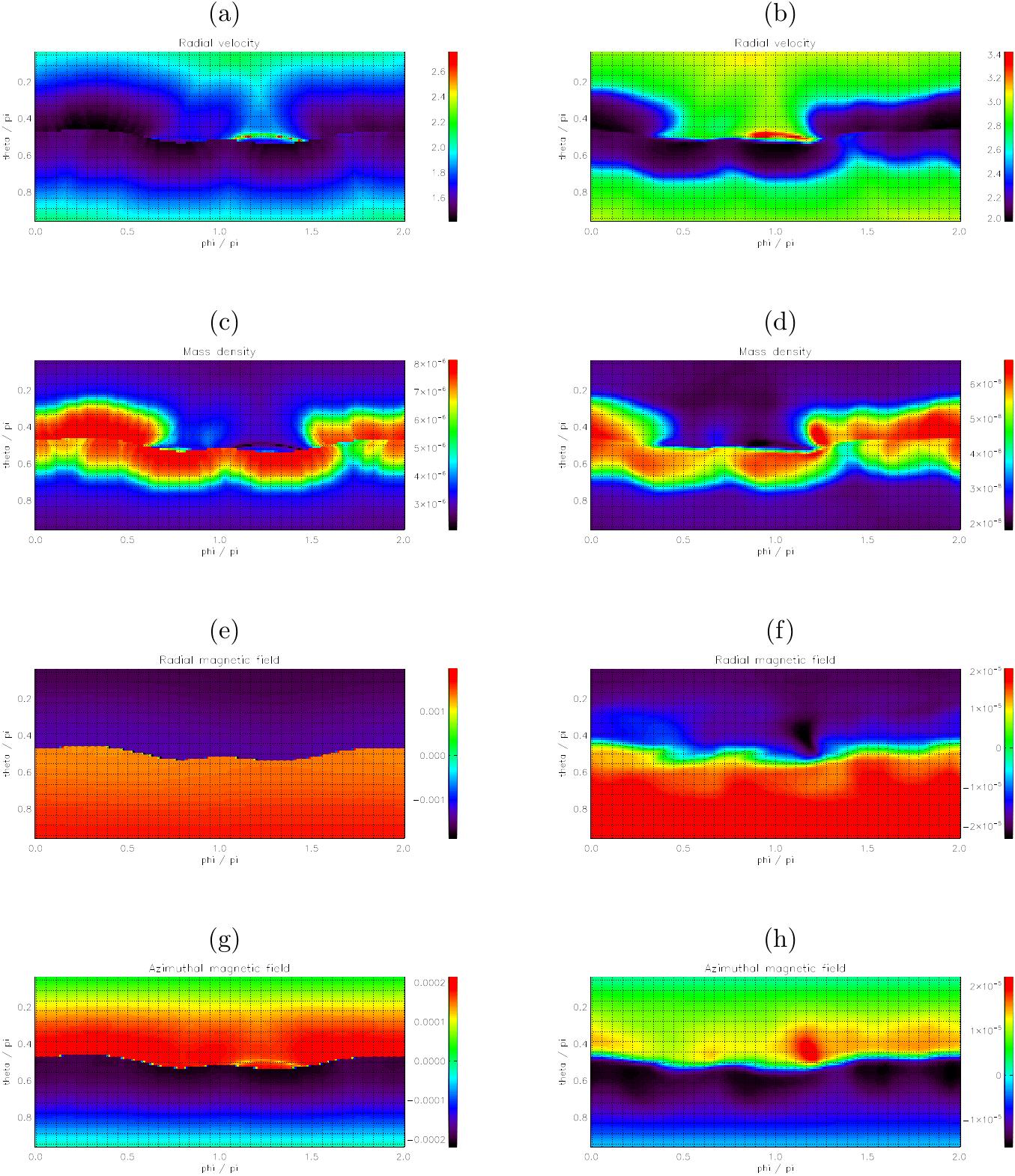}
\caption{\it Initial values at the inner radial grid boundary (left panels) and results at the outer radial grid boundary for the converged state (right panels) for solar minimum. The shown quantities are (top to bottom) radial velocity, mass density, radial- and azimuthal magnetic field. For the nomalizations see Table \ref{t:norm}.}
\label{fig:min}
\end{figure*}

\begin{figure*}
\noindent
\includegraphics[width=\textwidth]{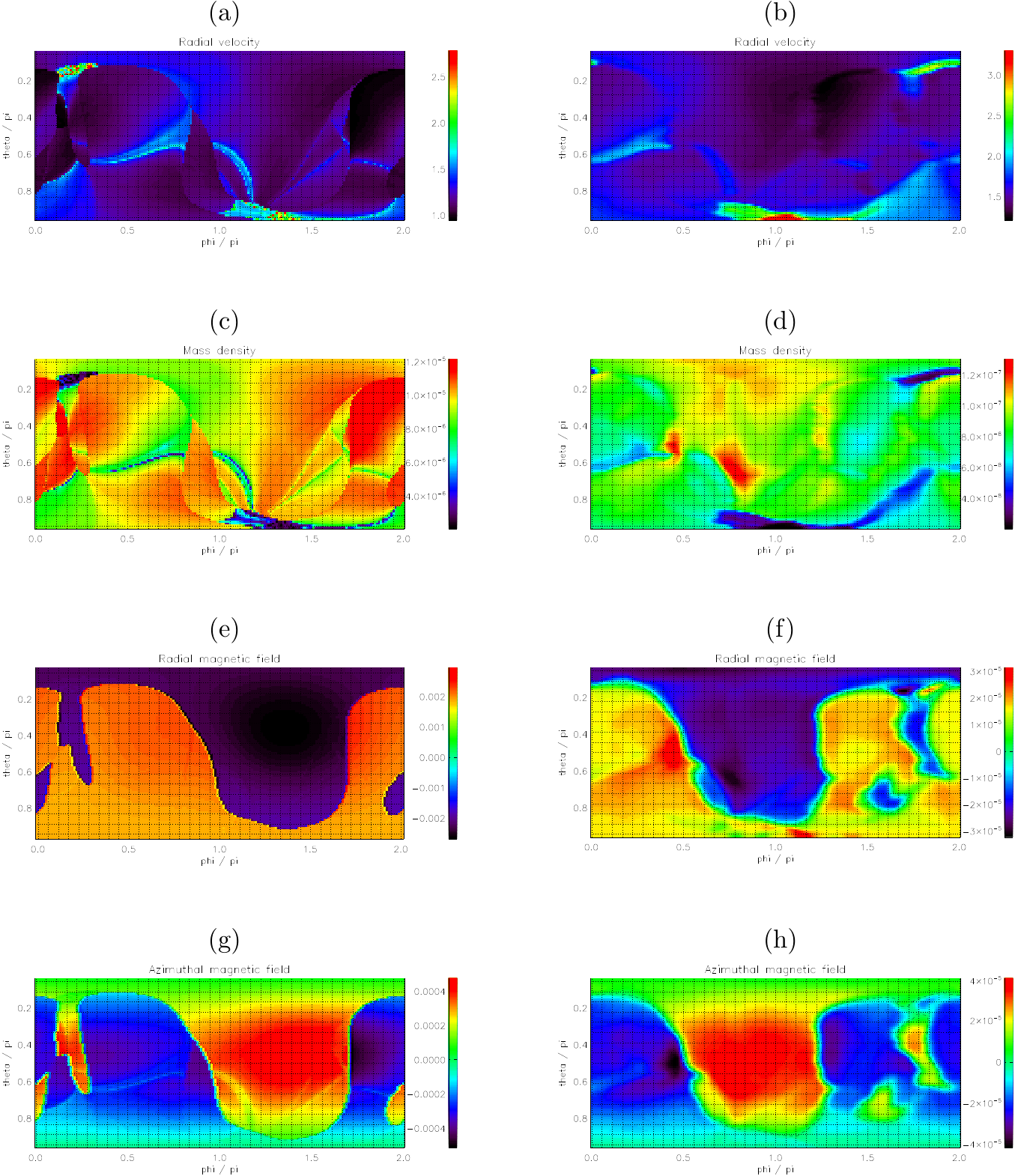}
\caption{\it Same as Figure \ref{fig:min} for solar maximum conditions.}
\label{fig:max}
\end{figure*}

Equation (\ref{det:vr2}) is based on the empirical inverse correlation between flux tube expansion factor $f_s$ (hereafter EF) and radial solar wind speed $v_r$ observed near Earth reported for potential field models as described in \citet{Wang-Sheeley-1990} and \citet{Wang-etal-1997}. Additionally, $d_{FP}$ is the footpoint distance to the nearest coronal hole boundary. The remaining unknowns are tuning parameters. The formula for the expansion factor reads
\begin{equation}
f_s = \left(\frac{R_\odot}{R_{ss}}\right)^2\cdot\frac{B(R_\odot,\vartheta_0,\varphi_0)}{B(R_{ss},\vartheta,\varphi)}
\end{equation}
where $B(R_\odot,\vartheta_0,\varphi_0)$ and $B(R_{ss},\vartheta,\varphi)$ are the magnetic field strengths at a point in the photosphere and (following a magnetic field line) at the source surface at $R_{ss}$. In order to know how points on the two respective magnetic field maps are connected, it is necessary to find a mapping $(\vartheta,\varphi)\mapsto(\vartheta_0,\varphi_0)$. The resulting footpoint locations are shown in Figure \ref{fig:footpoints} as black/red dots. During solar minimum (a) field lines originate from polar regions, but in this case there also exist excursions extending towards the equator. At solar maximum (b) there is no discernible structure to be made out: Field lines originate from coronal holes that can be located anywhere on the solar surface. To test whether these results are realistic, they can be compared to their observational counterparts, which are coronal holes inferred from National Solar Observatory He maps. This is shown for the case of solar maximum in Figure \ref{fig:footpoints}: Model data is shown in red dots, observational data in black contours. There is some discrepancy, because the model is based on very different data (sunspot areas) and is a snapshot, while the observed synoptic map is pasted together from 27 days of data (because the whole solar surface cannot be seen from Earth at a time). Additionally, the observational technique misses very thin structures. Keeping this in mind the match between the data sets seems reasonable.\\
The location of the gridpoints of the map of magnetic field strength at the solar surface do not necessarily coincide with the derived footpoint locations and interpolation is applied, allowing to compute the magnetic field strength at the footpoints of the open field lines $B(R_\odot,\vartheta_0,\varphi_0)$ and from this the expansion factor $f_s$ at every grid point on the source surface. The footpoint locations also allow to compute $d_{FP}$ for every corresponding grid point on the source surface by finding the nearest point that can be designated as part of a coronal hole boundary (yellow dots in Figure \ref{fig:footpoints}). Tuning parameters have been adopted from the original paper ($v_{min}=154$~km/s, $v_{del}=300$~km/s, $V_{exp} = 0.3$, $\alpha=7.4$~km/s, and $\beta=3.5$). The resulting initial radial velocities at the inner radial grid boundary are shown in the respective top left panels of Figures \ref{fig:min} and \ref{fig:max} for both minimum and maximum conditions.\\
Typical solar minimum conditions are rather well known (high-speed streams in polar regions and low-speed streams in the ecliptic) and this general structure is reconstructed here. An interesting feature of high-speed wind in the equatorial region is also found, and a comparison with OMNIweb {\tt http://omniweb.gsfc.nasa.gov/} data at the time in question does indeed show large equatorial speeds at the Earth at the time in question (see Figure \ref{fig:omni1987} showing wind speeds of up to 700~km/s), which supports the assumption that this feature is real. 
For solar maximum there is neither a typical structure nor a sufficient amount of out-of-ecliptic measurements to compare with. There are however attempts to describe solar maximum conditions \citep{Zurbuchen-etal-2004}, but a comparison is still difficult, because the background field is constantly distorted by the large number ($\propto5$/day) of CMEs (as is also pointed out in \citet{Zurbuchen-etal-2004}). We therefore omit a detailed comparison for solar maximum because CMEs are not incorporated in this model as of yet.  \\
Equation (\ref{det:rho}) uses the value of the radial velocity and in its original form assumes constant mass flux $F_{mass}$, thus yielding an inverse relation between mass density $\rho$ and radial velocity $v_r$. We calculated $F_{mass,0}$ by taking values of $\rho v_r$ at the Earth at 1~AU and applying an $r^2$ scaling to get values at 0.1~AU. For this, 27-day averages of both quantities at the time in question are taken from the OMNIweb interface. The adopted values can be found in Table \ref{t:massflux}.\\
\begin{table}
 \caption{\it 27-day averaged values for mass density $\rho$ and radial velocity $v_r$ at 1~AU taken from the OMNIweb interface.}
\begin{center}
   \begin{tabular}{ c | c | c  }
      year & $\rho$ / (cm$^{-3}$) & $v_r$ / (km/s) \\ \hline
     1987.2 & 6.0 & 508 \\ \hline
     2000.5 & 6.4 & 476 \\   
   \end{tabular}
 \end{center}
 \label{t:massflux}
\end{table}
We introduced an improvement to be in agreement with \it Ulysses \rm data \citep{McComas-etal-2000}, which showed for solar minimum conditions that the mass flux associated with high-speed streams is about half the one of low-speed streams. This is implemented by having the mass flux depend on the radial velocity by introducing a transition function $f(v_r)$ (see \citet{Scherer-etal-2010}) defined as
\begin{equation}
f(v_r) = \frac{1}{2}(\tanh(\Delta(v_r + v_t)) - \tanh(\Delta(v_r - v_t)))
\end{equation}
that gives a sharp, but steady transition at $v_t$ with $\Delta=20$. The value of $v_t$ is chosen such that the transition from high to low speed occurs at $\vartheta\approx\pm 35°$ for our solar minimum data, which yields $v_t = 1.5v_0$. Mass flux can then be calculated via
\begin{equation}
F_{mass} = (0.5 + 0.5f(v_r))F_{mass,0}~.
\end{equation}   
Results are  shown in Figures \ref{fig:min} and \ref{fig:max} (b).\\
Equation (\ref{det:T}) is the pressure balance between gas pressure and magnetic pressure solved for proton temperature (which is the only temperature in this one-fluid model). The tuneable parameter $p_{tot}$ is the total pressure, but its optimal values derived by \citet{Detman-etal-2006} change over the solar cycle, so that a rough tuning has been performed on our part to get realistic results and we chose $p_{tot,min}=3.8p_0$ and $p_{tot,max}=4.2p_0$ for solar minimum and maximum, respectively.\\

The radial magnetic field strength has to be scaled from source surface values $B_{ss}(R_{ss}$) to values at the lower radial grid boundary at $R_{gb}$. The scale factor simply mimics the usual $r^{-2}$ behavior, so that $b_{scale} = (R_{ss} / R_{gb})^2$. This is a very simple approach, which is not optimal everywhere, because an inhomogeneous azimuthal magnetic field component yields deviations from the $r^{-2}$ behavior for the radial magnetic field because of the solenoidality condition. However, at small radial distances, such deviations should not be too serious since azimuthal components are small.\\

Equations (\ref{det:vtet}) and (\ref{det:Btet}) set the $\vartheta$ components of both the velocity and the magnetic field equal to zero. We found that both $v_\vartheta$ and $B_\vartheta$ remain negligible throughout the computational volume, so that these approximations are affirmed.\\ 

As was demonstrated in the WD test case, the Parker spiral can be realized in two frames of reference (i.e.~one co-rotating with the Sun and the other at rest with respect to the Sun's rotation), both of which have been found to give equivalent results. Equations (\ref{det:vphi}) and (\ref{det:Bphi}) model the Parker spiral in the co-rotating frame, whereas in \citet{Detman-etal-2006} the rest frame is implemented. The equation for the azimuthal velocity mimics the behavior of the low corona rotating approximately like a rigid body, thus using the solar rotation frequency $\Omega$ multiplied with the radius of co-rotation, which \citet{Detman-etal-2006} put at $R_\varphi = 1.5R_\odot$. This approach is questionable in view of the behavior of the azimuthal velocity derived by \citet{Weber-Davis-1967} shown in Figure \ref{fig:WD}, because from the radius of co-rotation outwards an approximate $r^{-1}$ behavior is evident that would yield considerably smaller values at $R_{gb}$. Correction for this with a respective scale factor has been applied here, so that $R_\varphi = 1.5R_\odot/(R_{gb}/R_\odot)$. This has been done in the recent update in \citet{Detman-etal-2011} as well, as $R_\varphi$ is now described as "a length scale which determines the azimuthal velocity", but no specific value is given to compare with. This is still a rather crude estimate since the radius of co-rotation depends on the magnetic field strength near the Sun and simple $1/r$ scaling does not exactly reproduce the azimuthal velocity solution that should be similar to the WD solution. But for this simple empirical model it shall suffice.\\
The equation for the azimuthal magnetic field (\ref{det:Bphi}) is the result of the induction equation in the rotating steady state corona in corotating spherical coordinates. This is a different approach then the one used for the testcases (see Equation (\ref{eq:bphi_test})), which became necessary for the following reason: In the Weber-Davis-model the azimuthal magnetic field is homogeneous and can be extrapolated into the inner boundaries ghost cells from the values obtained from the innermost computed cells. The observationally based input for the magnetic field used here, however, is inherently inhomogeneous and makes extrapolation a challenging task, because of the staggered mesh implemented in CRONOS. Occasionally, this even gave rise to instabilities. To avoid this, we chose to fix the azimuthal magnetic field components according to Equation (\ref{det:Bphi}). In view of the inherently empirical nature of the whole set of equations to determine the remaining MHD quantities, this approach seems justifiable.

\subsection{Radial initialisation}
The values of the MHD quantities at the lower radial boundary derived above will now also serve to find initial values for the rest of the numerical domain. Here, radial velocity along with mass density is to fulfill the stationary equation of continuity, which is achieved by having $v_r\propto r$ and $\rho\propto r^{-3}$. Azimuthal velocity in the co-rotating frame of reference is basically proportinal to radial distance ($v_\varphi\propto r$). For adiabatic expansion the radial dependence of temperature should approximately behave like $T\propto r^{-4/f}$ where $f$ is the number of degrees of freedom, related to the polytropic index via $f=2/(\gamma-1)$. In our case of $\gamma=1.05$ this gives $f=40$.\\
The magnetic field has to be initialized in a divergence-free manner, so that for later times the code retains a divergence-free field due to the constrained transport algorithm. This has been achieved by going through every cell layer radially, and solving the solenoidality condition to find the next cells magnetic field components. A detailed description of this algorithm would require a discussion of the grid layout and the operation of the code, and is therefore omitted here \citep[, but see][]{Kissmann-etal-2009,Kissmann-Pomoell-2012}.

\subsection{Boundary conditions \& Grid layout}
\label{bc2}
Boundary conditions are chosen similarly to those used for the test cases, i.e.,~reflecting $\vartheta$ boundaries, periodic boundary conditions for $\varphi$, and an outflow condition at the outer radial boundary. At the inner radial boundary specific conditions are applied for each on quantity.\\
The radial velocity and mass density were kept fixed at the inner radial boundary. Additionally, the ghost cells are assigned values derived from extrapolation according to the governing power law in the first computed cell. Magnetic field components and $v_\vartheta$ remain fixed to their initial values as well, but no extrapolation for the ghost cells is applied. The azimuthal velocity $v_\varphi$ is expected to be dominated by the transformation term to the co-rotating frame of reference, giving $v_\varphi\approx-\Omega r\sin(\vartheta)$. Thus, a radial extrapolation for the inner radial ghost cells is applied according to 
\begin{equation}
v_{\varphi,i} = \frac{r_i}{r_0}v_{\varphi,0}
\end{equation}
where $i$ indicates the negative index of the ghost cells, while the first cell in the computational domain carries the index 0.\\
For the observationally based input data the following grid layout has been chosen:
For the latitudinal coordinate the ghost cells at each boundary should not coincide with or go beyond the poles since in the current setup coordinate singularities have to be excluded, but this can be treated in the near future as well. Here, we are not primarily concerned with the polar regions.\\
Furthermore, using the highest possible angular resolutions ($1°\times1°$) with an adequate radial resolution was found to require too long to converge. Instead we took half the possible angular resolution and a comparison with full resolution simulations yielded no significant differences. This gives $\vartheta_{grid}\in[8°,172°]$ and a cell size of $\Delta\vartheta=2°$ gives 172 cells. Similarly, for the longitudinal coordinate $\varphi_{grid}\in[0°,360°]$, requiring 180 cells. The radial length of the simulation box is determined by Earth's orbital distance ($215R_\odot$) and the lower radial grid boundary (at $21.5R_\odot$), where we chose a radial cell size of $1.5R_\odot$.\\
\\
\\
\\
 
\section{Simulation results}
\subsection{Global magnetic field structure}
\label{results}
\begin{figure*}
\begin{center}
\includegraphics[width=\textwidth]{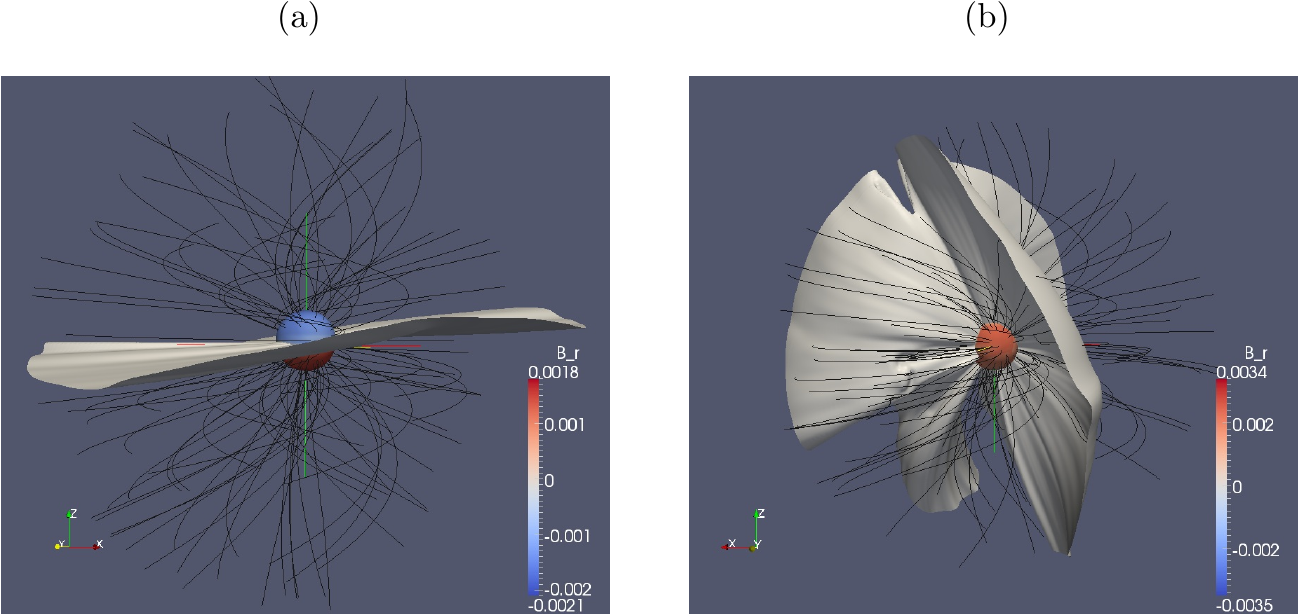}
\end{center}
\caption{\it Visualization of magnetic field topologies from the start surface out to 1~AU at solar minimum (a) and maximum (b): The spherical inner radial grid boundary is color-coded with the radial magnetic field $B_r$, and the current sheet is the contour of $B_r=0$. Field lines are shown in black and demonstrate the bending caused by solar rotation.}
\label{fig:paraview}
\end{figure*}
To get an impression of the results, pseudo three-dimensional visualizations of the respective data sets are shown in Figure~\ref{fig:paraview}. The inner radial grid boundary can be extrapolated from the set of data by performing a spherical clip at the respective radius (giving the sphere in the center). After applying $1/r^2$ scaling, the colorbar is for the radial magnetic field strength $B_r$, and is in accordance with the input data as was shown in Figure \ref{fig:Rss}. The current sheet is the $B_r=0$ contour through the computational grid. It can be seen that, at solar minimum, the current sheet is slightly wavy as expected from the input data, and it is subject to solar rotation as well, which can be seen by the bent wavy features. Magnetic field lines are colored black and are not uniformly distributed. The spiral structure can be made out as well by the bent field lines. For solar maximum conditions the current sheet's structure is rather extreme, with steep gradients in excursions towards the poles.\\

\subsection{Plasma structure at 1~AU}
\begin{figure*}
\includegraphics[width=\textwidth]{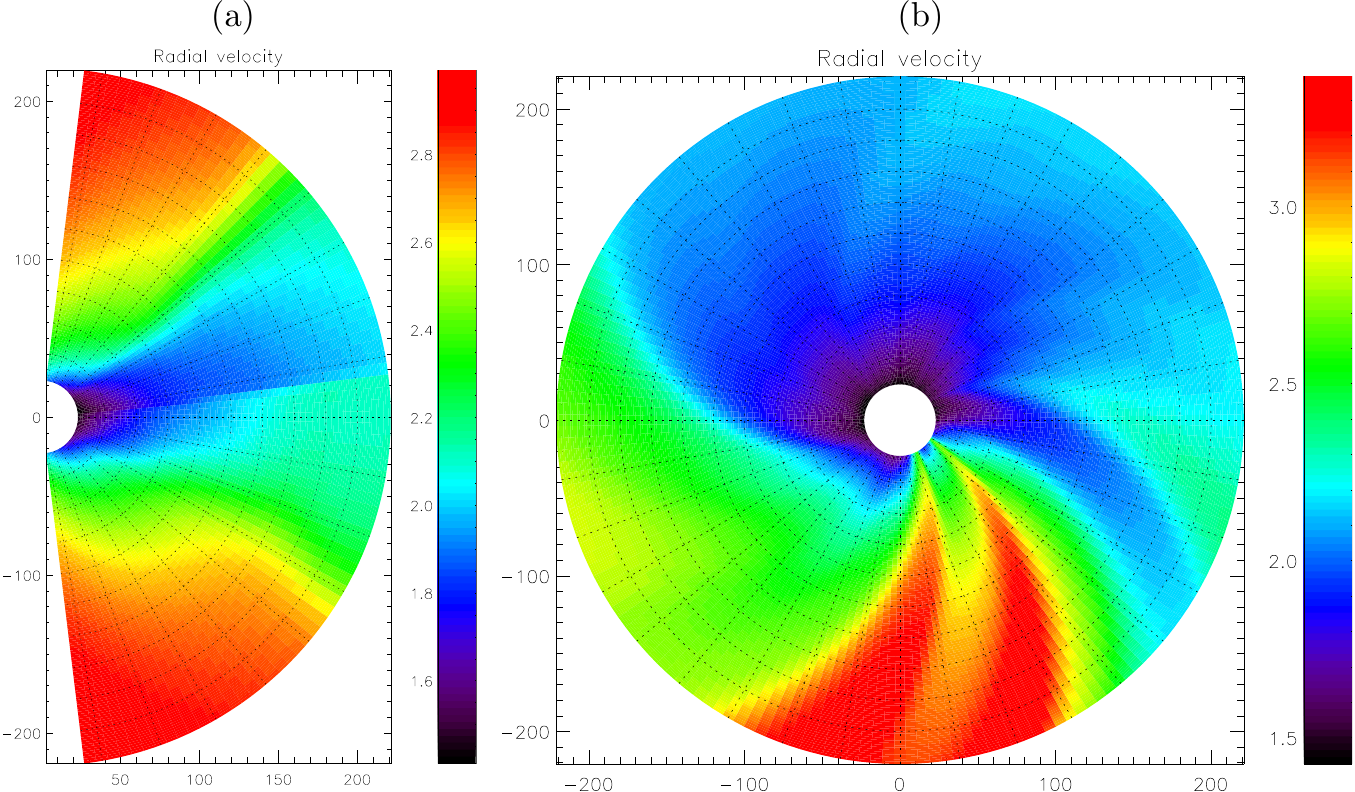}
\caption{\it (a) Meridional ($\varphi=0$) and (b) equatorial slice through the computational grid for radial velocity.}
\label{fig:vrad}
\end{figure*}
The results at the outer radial grid boundary for radial velocity, mass density and magnetic field components are shown on the right hand sides of Figures \ref{fig:min} and \ref{fig:max}, respectively. \\
Comparing with the initial data, it can be seen that the results are strongly dependent on the input data, i.e.~the input determines the results. Initial features can still be observed at the outer radial boundary but appear smoother. The effect of solar rotation is also visible as features shift in longitude. At solar minimum, for velocity and mass density it can be observed that the transit from slow to fast speed is sharper than it was initially due to the differently strong acceleration. In equatorial regions we get speeds of about 400~km/s and the high-speed feature is at 700~km/s in accordance with the observational data taken from the OMNIweb (Figure \ref{fig:omni1987}). At polar regions, speeds of about 650~km/s are found. Since there is no observational data for polar regions from that time it cannot be judged whether this value agrees with the conditions present at that time. 
Mass density is inversely correlated with radial velocity and the obtained values at Earth's orbit again agree with the observational data. The features of the magnetic field input are also preserved but the current sheet slightly broadens, which may be due to the finite spatial resolution. An interesting feature that is not present initially occurs at the location of the high-speed feature where we find an enhancement of the magnetic field strength. This is associated with the compression of the magnetic field when fast solar wind runs into slow one ahead, also known as co-rotating interaction regions.\\
At solar maximum there is no typical global structure in wind speed features to compare with so that with the limited number of observational data it is difficult to determine the quality of our results. It can, however, be stated (two top-right panels of Figure \ref{fig:max}) that the obtained values for wind speed and mass density are again in the right order compared to the observational data. The initial magnetic field is largely preserved but is again modulated by features in the wind speed. 
 
\subsection{3-D velocity structure}
\begin{figure}
\noindent
\begin{center}
\includegraphics[width=0.45\textwidth]{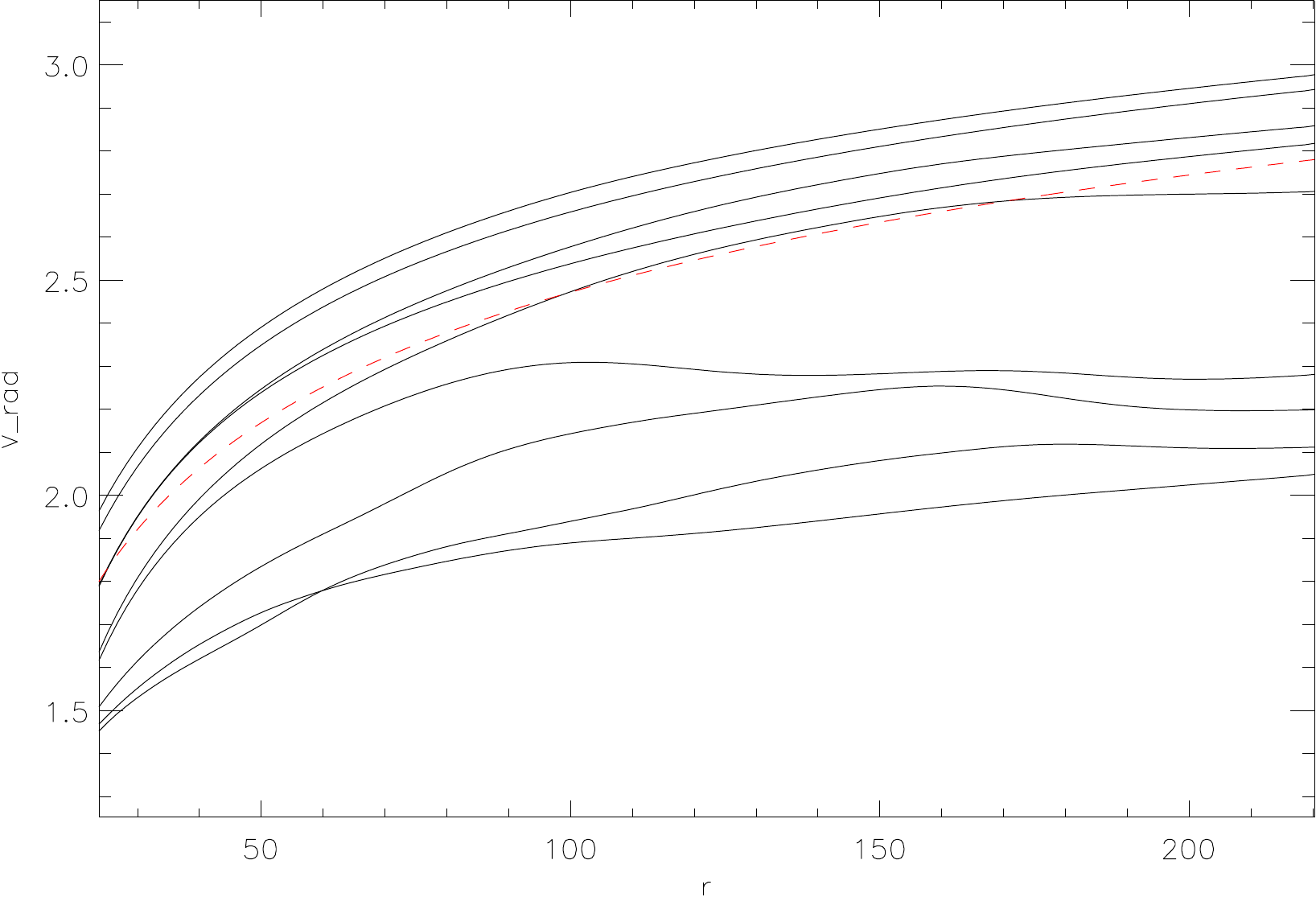}
\end{center}
\caption{\it Radial velocity profiles at different latitudes accompanying the meridional slice of Figure \ref{fig:vrad}. The top curves correspond to high-speed polar regions and the bottom ones to equatorial regions. The red-dashed line is the Parker solution for a temperature $T=0.49T_0=1.4\cdot10^6{\rm K}$ corresponding to the initial temperatures at polar regions, and shows that our results are close to it.}
\label{fig:vrad2}
\end{figure}

The radial velocity profiles at solar minimum are shown in Figure \ref{fig:vrad} in a meridional ($\varphi=0$) and an equatorial ($\vartheta=\pi/2$) slice through the computational grid. The meridional slice illustrates the dependence on latitude and Figure \ref{fig:vrad2} shows a line plot accompanying the meridional slice, in which the radial velocity curves for different latitudes can be seen. The top curves correspond to high-speed polar regions and the bottom ones to equatorial regions. The red-dashed line is the Parker solution for a temperature $T=0.49T_0=1.4\cdot10^6{\rm K}$ corresponding to the initial temperatures at polar regions, and shows that our results are close to it. The drop in some of the curves is due to slow wind features shifting to this longitude at larger radial distances, which is an effect of solar rotation. These effects are also visible in the right panel of Figure \ref{fig:vrad}. At close inspection it can also be seen that the shift in longitude between the start surface and the outer radial grid boundary due to solar rotation depends on the radial velocity. This corresponds to the so-called Parker angle that is the ratio between azimuthal and radial components. At the Earth it is usually at about 45° for radial velocities of about 400~km/s, which is also the case here near the ecliptic and we, too, find the corresponding shift in longitude to be at about 45°. We compared the winding angle (i.e.~the ratio between the magnetic field components) with the Parker prediction and found good agreement for all latitudes except at the current sheet, which is understandable, because the Parker prediction does not involve dipolar fields.

\section{Conclusions and outlook}
In this work we successfully used observationally based input data of the radial magnetic flux to drive an inner heliospheric MHD simulation to calculate solar wind quantities at radial distances up to 1~AU. Testing the code has been done by a comparison to the simplified analytical models by \citet{Parker-1958} and \citet{Weber-Davis-1967}. It could be demonstrated that the code succesfully reproduced the analytic results.\\
The input data for the radial magnetic flux and its derivation are described in \citet{Jiang-etal-2010}, which amongst several advancements allows for higher angular resolution than that of comparable observational data from synoptic magnetograms. While here the focus lay on two representative sets of data, those being one set at solar minimum and one at solar maximum, data are available for a whole time series going back to 1976. The possibility to run the code for more sets of data is present, including the opportunity to combine them to drive a time-dependent simulation.\\
The MHD code requires input data for all plasma quantities, and it is necessary to use empirical formulas to derive them. The respective set of equations was taken from \citet{Detman-etal-2006, Detman-etal-2011}, who work on a very similar goal, i.e.~space weather forecasting, for which they use input data similar to those used here. These kinds of interfaces are constantly augmented as testing goes on. In this respect, this work can be understood to be part of this process with the difference of using more sophisticated input data and applying the empirical formulas developed and tuned for a specific code to the one used here.\\
The results give values for solar wind quantities at 1~AU that can be used as input for outer-heliospheric large-scale models \citep[e.g.][]{Ferreira-etal-2007a,Scherer-etal-2008a}, as background solar wind for coronal mass ejection simulations \citep[e.g.][]{Kleimann-etal-2009} or to model solor wind plasma structures \citep[e.g.][and references therein]{Dalakishvili-etal-2011} such as co-rotating interaction regions.

%
%
%
%
%
%
%

\begin{acknowledgments}
Financial support for the project FI 706/8-2 (within the Research Unit 1048) and for the project FI 706/14-1 funded by the Deutsche Forschungsgemeinschaft(DFG) is acknowledged.
\end{acknowledgments}

%
%
%
%
%
%
%
%
%


%
%

\end{article}


%
%

%
%
%
%
%
%
%


\end{document}